 \documentclass[final,3p,times]{elsarticle}
\usepackage{amsmath,graphicx,amssymb,color,cancel,float}
\usepackage{color}

\newcommand{\be}{\begin{eqnarray}}
\newcommand{\ee}{\end{eqnarray}}

\journal{Physics Letter B}

\begin{document}

\begin{frontmatter}

%% Title, authors and addresses

%% use the tnoteref command within \title for footnotes;
%% use the tnotetext command for theassociated footnote;
%% use the fnref command within \author or \address for footnotes;
%% use the fntext command for theassociated footnote;
%% use the corref command within \author for corresponding author footnotes;
%% use the cortext command for theassociated footnote;
%% use the ead command for the email address,
%% and the form \ead[url] for the home page:
%% \title{Title\tnoteref{label1}}
%% \tnotetext[label1]{}
%% \author{Name\corref{cor1}\fnref{label2}}
%% \ead{email address}
%% \ead[url]{home page}
%% \fntext[label2]{}
%% \cortext[cor1]{}
%% \address{Address\fnref{label3}}
%% \fntext[label3]{}

\title{Quantum Electroweak Symmetry Breaking \\ Through Loop Quadratic Contributions}

%% use optional labels to link authors explicitly to addresses:
%% \author[label1,label2]{}
%% \address[label1]{}
%% \address[label2]{}

\author{Dong Bai}
\ead{dbai@itp.ac.cn}
\address{State Key Laboratory of Theoretical Physics(SKLTP), Kavli Institute for Theoretical Physics China (KITPC)\\ Institute of Theoretical Physics,
Chinese Academy of Sciences, Beijing, 100190, China \\
University of Chinese Academy of Sciences (UCAS), Beijing, 100049, China }

\author {Jian-Wei Cui}
\ead{jwcui@hust.edu.cn}
\address{School of Physics, Huazhong University of Science and Technology, 1037 Luoyu Road, Wuhan, China} 

\author{Yue-Liang Wu \corref{ylwu}} 
\ead{ylwu@itp.ac.cn}  
\address{State Key Laboratory of Theoretical Physics(SKLTP),
Kavli Institute for Theoretical Physics China (KITPC)\\
Institute of Theoretical Physics,
Chinese Academy of Sciences, Beijing, 100190, China\\
University of Chinese Academy of Sciences (UCAS), Beijing, 100049, China} 

\cortext[ylwu]{Corresponding author}

\begin{abstract}
%% Text of abstract
Based on two postulations that (i) the Higgs boson has a large bare mass $m_H \gg m_h \simeq 125 $ GeV  at the characteristic energy scale $M_c$ which defines the standard model (SM) in the ultraviolet region, and (ii) quadratic contributions of Feynman loop diagrams in quantum field theories are physically meaningful, we show that the SM electroweak symmetry breaking is induced by the quadratic contributions from loop effects. As the quadratic running of Higgs mass parameter leads to an additive renormalization, which distinguishes from the logarithmic running with a multiplicative renormalization, the symmetry breaking occurs once the sliding energy scale $\mu$ moves from $M_c$ down to a transition scale $\mu =\Lambda_{EW}$ at which the additive renormalized Higgs mass parameter $m^2_H(M_c/\mu)$  gets to change the sign. With the input of current experimental data, this symmetry breaking energy scale is found to be $\Lambda_{EW}\simeq 760$ GeV, which provides another basic energy scale for the SM besides $M_c$.  Studying such a symmetry breaking mechanism could play an important role in understanding both the hierarchy problem and  naturalness problem. It also provides a possible way to explore the experimental implications of the quadratic contributions as $\Lambda_{EW}$  lies within the probing reach of the LHC and the future Great Collider.
\end{abstract}

\begin{keyword}
%% keywords here, in the form: keyword \sep keyword
quadratic contribution \sep loop regularization \sep electroweak symmetry breaking
 
%% PACS codes here, in the form: \PACS code \sep code

%% MSC codes here, in the form: \MSC code \sep code
%% or \MSC[2008] code \sep code (2000 is the default)

\end{keyword}

\end{frontmatter}

%% \linenumbers

%% main text

The advent of the Standard Model (SM)-like Higgs boson at the LHC \cite{higgs2012a,higgs2012c} initiates us to investigate in detail the properties of the SM Higgs sector and to understand deeply the electroweak symmetry breaking mechanism. It is known that the most distinguishing feature of the SM Higgs sector is the existence of quadratic contributions (or quadratic divergences) in the loop calculations of Feynman diagrams. It involves the long-lasting issues on the so-called hierarchy problem and naturalness problem. The main questions include whether the Higgs boson is a fundamental particle or a composite one, and what  is the energy scale for a possible new physics beyond the SM. As it is well understood that the gauge couplings of interactions and masses of particles are all running quantities of the energy scale. The well-known fact is the discovery of asymptotic freedom of gauge coupling in QCD \cite{Gross:1973id,Politzer:1973fx}, which has led to the motivation of grand unification theory for all gauge interactions. The running behavior of all physical quantities in the SM is logarithmic except for the Higgs mass parameter. For the logarithmic running, it is known to result in a multiplicative renormalization. As a consequence, such a running is always proportional to the coupling/mass itself. For the Higgs mass parameter, if ignoring the quadratic contributions, one yields a similar behavior.  When taking into account the quadratic contributions to the Higgs mass parameter, one yields a quadratic running for the Higgs mass parameter. Unlike the logarithmic running, the quadratic running will result in an additive renormalization. In fact, it was shown that the quantum gravitational contributions can also cause a quadratic running to gauge couplings \cite{RW,TW}. How to treat and understand the quadratic running of the Higgs mass parameter comes to the main issue in this paper. 

It has been known for a long time that quantum field theory (QFT) is ``plagued" by ultraviolet (UV) divergences. In the modern viewpoint based on Wilsonian effective field theory \cite{wilson1974,polchinski1984}, QFT is typically defined with respect to some physical UV cutoffs, and the UV divergences are reinterpreted as extra contributions from the UV modes to the parameters in the low-energy effective Lagrangians. In this sense, we shall use the words ``UV contribution" rather than ``UV divergence" in the quantum loop calculations. In the Higgs sector of the SM,  there is a long-established implication that the quadratic contributions, unlike the logarithmic contributions, can lead to unwanted over-contributions from the UV modes to the low-energy Higgs mass parameter, unless there exist some extremely delicate tunings between these quadratic contributions and the so-called bare Higgs mass parameter.  Therefore, to achieve a natural explanation for the observed Higgs mass rather than asking the huge fine-tunings, it is important to treat properly these quadratic contributions. This is often called the naturalness problem in literature. Meanwhile, if there is no new physics above the electroweak scale up to the grand unification scale or Planck scale, it then raises the so-called hierarchy problem. In the last three decades, these problems have drawn enormous attentions. To solve these problems, one chooses either to eliminate these quadratic contributions in the UV regions or to lower the ultimate scale from the Planck scale to the TeV scale. Another solution is based on the consideration that the discovered Higgs is not a fundamental scalar particle, so that the calculations for the quadratic contributions cannot be extended to scales much higher than the characteristic energy scale of the composite Higgs models.  Alternatively, one even argues that the naturalness problem is not a right question at all, and nature just behaves in an unnatural way. Popular studies along these directions include the electroweak scale supersymmetry \cite{martin2011}, extra dimension models \cite{cheng2010}, composite Higgs models \cite{contino2010}, multiverse scenario \cite{arkani-hamed2005}. Notably, almost all the solutions (with the multiverse scenario as a representative counterexample) can lead to new physics at the TeV scale. In all the considerations, the quadratic contributions are treated to be the inevitable loop quantum effects in QFTs and have to be tamed carefully to avoid the intolerable unnaturalness. 

On the other hand, it has been demonstrated that in the effective field theories, such as the chiral dynamical model of QCD, the quadratic contributions play a crucial role for the derivation of the gap equation to describe the dynamically generated spontaneous chiral symmetry breaking \cite{gherghetta1994,daiwu2004}, where the scalar mesons have been regarded as the composite Higgs-type bosons \cite{daiwu2004}. In the chiral perturbation theory, the quadratic contributions have been shown to be significant in understanding the $\Delta I=1/2$ selection rule on the isospin amplitudes $A_0/A_2$ \cite{BBG1987} and the direct CP-violating parameter $\epsilon ' /\epsilon$ \cite{HPSW1992}, and to provide simultaneously a consistent explanation for both the ratio $\epsilon ' /\epsilon$ and the isospin amplitudes $A_0/A_2$ in the Kaon decays \cite{wu2001}. Recently,  it was noticed in Ref.~\cite{hamada2012} that the coefficient of the quadratic contributions of the SM Higgs sector has a novel zero point around the scale of $10^{23}\ \text{GeV}$,  which provides new insights into the Higgs inflation scenario \cite{hamada2013,hamada2014} and also the possible hierarchy problem solution \cite{chankowski2014} based on the idea of softly broken conformal symmetry \cite{bardeen1995}. 

Theoretically, in the perturbative expansions and calculations of the SM, there exist in general both quadratic and logarithmic divergences. When adopting the dimensional regularization scheme, the quadratic divergences are suppressed due to the analytical extension for the space-time dimensions of original theories. Although the dimensional regularization scheme is practically very convenient in calculations and widely recognized in literature, it is inevitable to result in divergences when taking the exact space-time dimensions to recover the original theories. Thus the dimensional regularization scheme is actually in spirit incompatible with the modern framework of effective field theories, especially when the theories involve quadratic contributions. It is natural to find out a new scheme which is suitable for the modern framework of effective field theories and applicable for all QFTs while preserving symmetry properties and divergent structures of original theories. It has turned out that the loop regularization (LORE) method proposed in Ref.~\cite{wu2003} is a concrete realization for such new schemes. It has been shown explicitly at the one-loop level that the LORE method can preserve Poincare symmetry and gauge symmetry \cite{wu2003}, even supersymmetry \cite{cuitangwu2009}, and can be extended consistently beyond the one loop \cite{huangwu2012,HLW2013}. Unlike the dimensional regularization scheme, the LORE method is realized in the exact space-time dimensions of original theories and all the calculations can be done exactly without modifying original theories. All the UV contributions, both quadratic and logarithmic, in the Feynman integrals can be calculated in a unified manner as the divergent integrals can well be defined in the LORE method. In particular, the LORE method is found to be an infinite-free regularization scheme which leads to finite results characterized by two intrinsic energy scales. These two intrinsic energy scales are introduced naturally in the LORE method to play the roles as the characteristic energy scale $M_c$ and the sliding energy scale $\mu$, which can be identified as the UV energy scale to define the so-called bare Lagrangians and the infrared (IR) energy scale to yield the low energy effective Lagrangians respectively. As a consequence, the LORE method enables us to define a finite renormalization theory of QFTs. For more details on the LORE method, it is referred to the original papers and the recent review \cite{wu2014} for the interested readers. 

Based on the above mentioned considerations and motivated from the quadratic running of Higgs mass parameter and the null results of the new physics searching at the LHC Run I,  we are going to make two basic postulations: (i) the Higgs boson has a large bare mass $m_H \gg m_h\simeq 125$ GeV at the characteristic energy scale $M_c$ which defines the SM in the UV region, and (ii) the quadratic contributions of loop quantum effects are physically meaningful.  The first postulation implies that the Higgs sector should be different from the one in SM at the UV characteristic energy scale $M_c$. Let us begin with the following Lagrangian for the Higgs sector
\begin{equation}
\mathcal{L}_H=(D_{\mu}H)^{\dagger}(D^{\mu}H)-m_H^2H^{\dagger}H-\frac{\lambda_H}{4}(H^{\dagger}H)^2 -y_t\bar{Q}_Lt_R\tilde{H}+h.c., 
\end{equation}
where $m_H$, $\lambda_H$ and $y_t$ are the bare Higgs boson mass, Higgs coupling constant  and top-quark Yukawa coupling constant respectively, they all are defined at the UV characteristic energy scale $M_c$.  $H(x)$ is the Higgs doublet with $\tilde{H}=i\sigma_2H^* $. The covariant derivative $D_{\mu}$ is defined as 
\begin{equation}
D_{\mu}=\partial_{\mu}-ig_2I^a_WW^a_{\mu}+ig_1\frac{Y}{2}B_{\mu}, \nonumber
\end{equation}
where $g_2$ and $g_1$ are the gauge couplings for the $SU(2)_L$ and $U(1)_Y$ group respectively at the UV characteristic energy scale $M_c$. $I^a_W$ is related to the Pauli matrices via $I^a_W=\sigma^a/2$. Other interaction terms are the same as the ones in the SM with all the couplings defined at the UV characteristic energy scale $M_c$. 

In this article we will pay attention to the issue on the quadratic contributions to the above Higgs sector. We perform a calculation by using the LORE method and obtain a finite renormalized result for the UV contributions to the Higgs mass parameter. The low-energy effective Higgs mass parameter at the sliding energy scale $\mu$ is explicitly given by 
\begin{eqnarray} \label{QDHiggsMass}
m_H^2(M_c/\mu) & \simeq & m_H^2 \, ( 1 - \frac{3}{2(4\pi)^2} \lambda_H \text{ln}\frac{M_c^2}{\mu^2}  ) -\frac{6}{(4\pi)^2}\left(y^2_t-\frac{1}{4}\lambda_H-\frac{1}{8}g^2_1-\frac{3}{8}g^2_2\right)\left(M_c^2-\mu^2 \right) \nonumber \\
& = & \tilde{m}_{H}^2(M_c/\mu)  -\frac{6}{(4\pi)^2}\left(y^2_t-\frac{1}{4}\lambda_H-\frac{1}{8}g^2_1-\frac{3}{8}g^2_2\right)\left(M_c^2-\mu^2 \right) \, .
\end{eqnarray}
Here $\tilde{m}_{H}^2(M_c/\mu) $ is the logarithmic renormalized Higgs mass parameter  
\be
\tilde{m}_{H}^2(M_c/\mu) = m_H^2\, \left( 1 - \frac{3}{2(4\pi)^2} \lambda_H \text{ln}\frac{M_c^2}{\mu^2}  \right) 
\ee
where $m_H$ can be regarded as the Higgs mass parameter defined at the scale $\mu=M_c$, i.e., $m_H = m_H(M_c/\mu)|_{\mu=M_c}$. In obtaining above result, only the dominant top quark contribution is considered and the contributions from other quarks and leptons are ignored.  The above quadratic contributions were first calculated by Veltman \cite{veltman1981} at the one-loop level using the dimensional regularization. Where a trick has been used to extract the simple poles of the regularized Feynman integrals by taking the space-time dimension to be $D= 2$ rather than $D=4$, which is contrary to the calculation of the logarithmic divergences \cite{veltman1981}. Such a shift of poles often blurs the existence of quadratic divergences and becomes the main reason for the wrong impression that there is no quadratic divergence once the dimensional regularization is realized at $D= 4-\epsilon$ with $\epsilon \to 0$. The calculations at the two-loop level were carried out in Ref.~\cite{alsarhi1992}. Recently, a calculation was performed in Ref.~\cite{hamada2012} at the two-loop level and the same results were obtained. For our current purposes, it is useful to take the one-loop results for a demonstration on the electroweak symmetry breaking induced via the loop quadratic contributions.

It is noted that the characteristic energy scale $M_c$ denotes the UV scale which defines the SM, while the sliding energy scale $\mu$ is the infrared (IR) scale probed by the low-energy experiments. $m_H^2(M_c/\mu)$ is regarded as the additive renormalized Higgs mass parameter defined at the IR scale $\mu$. The combination of the coefficients $(y^2_t- \lambda_H/4-g^2_1/8-3g^2_2/8)$ takes the approximate value at the UV scale $M_c$. Rigorously speaking, Eq.~\eqref{QDHiggsMass} holds only when the truncated scale $\mu$ is numerically close to the UV scale $M_c$, so that the evolution of the coupling constants like $\lambda_H$, $g_1$, $g_2$ and $y_t$ and the logarithmic correction to the bare Higgs mass can be safely neglected. When the scale $\mu$ slides away far from the UV scale $M_c$, one has to adopt the renormalization group equations to obtain their evolutions. At the one-loop level,  the renormalization group equations (RGEs) are given by:
\be \label{QDHiggsMassRGE}
& & \mu\frac{d m_H^2}{d \mu}=\frac{12\mu^2}{(4\pi)^2} \bigg(y_t^2-\frac{1}{4}\lambda_H-\frac{1}{8}g_1^2-\frac{3}{8}g_2^2\bigg) + \frac{3\lambda_H}{(4\pi)^2} \tilde{m}_{H}^2, \\
& &  \mu\frac{d \tilde{m}_H^2}{d \mu} = \frac{3\lambda_H}{(4\pi)^2} \tilde{m}_{H}^2,
\ee
for the Higgs mass parameter, and 
\begin{eqnarray} \label{gaugecouplingRGE}
\mu\frac{d g_1}{d \mu}=\frac{g_1^3}{(4\pi)^2}\frac{41}{6}, \quad  \mu\frac{d g_2}{d \mu}=\frac{g_2^3}{(4\pi)^2}\left(-\frac{19}{6}\right), \quad \mu\frac{d g_3}{d \mu}=\frac{g_3^3}{(4\pi)^2}(-7),
\end{eqnarray}
\begin{equation} \label{topyukawaRGE}
\mu\frac{d y_t}{d \mu}=\frac{y_t}{(4\pi)^2}\bigg(\frac{9}{2}y_t^2-8g_3^2-\frac{9}{4}g_2^2-\frac{17}{12}g_1^2\bigg), 
\end{equation}
\begin{equation} \label{higgscouplingRGE}
\mu\frac{d \lambda_H}{d \mu}=\frac{2}{(4\pi)^2}\left[\lambda_H\left(3\lambda_H+6y_t^2-\frac{9}{2}g_2^2-\frac{3}{2}g_1^2\right)-12y_t^4+\frac{9}{4}g_2^4+\frac{3}{4}g_1^4+\frac{3}{2}g_2^2g_1^2\right],
\end{equation}
for the coupling constants. They hold for the sliding energy scale $\mu>\Lambda_{EW}$, where $\Lambda_{EW}$ is the electroweak symmetry breaking energy scale to be discussed below. Note that the strong interaction has no direct interactions to the Higgs field, but it does enter on the stage in the RGE approach through the $g_3$ dependence in the beta function of the top Yukawa coupling $y_t$. 

 It is seen from the above RGE Eq.(\ref{QDHiggsMassRGE}) that the quadratic running leads to an additive renormalization, which is different from the logarithmic running that gives a multiplying renormalization. Taking into account the above evolutions of the Higgs mass parameter and coupling constants as the energy scale $\mu$, we are going to demonstrate the implications of the quadratic contributions in the Higgs sector.  For our present purpose, it is more convenient to take the {\it integrated} expression for the additive renormalized Higgs mass parameter $m_H^2(M_c/\mu)$ defined  in Eq.~\eqref{QDHiggsMass} to show the properties of the loop quadratic contributions. For a numerical calculation, one shall take the RGE approach to obtain more accurate quantitive results via Eq.~\eqref{QDHiggsMassRGE}. It is interesting to observe that when the coefficient associated to the quadratic contributions is positive, i.e., 
\be
y_H^2(M_c/\mu) \equiv y^2_t-\lambda_H/4-g^2_1/8-3g^2_2/8 > 0,
\ee
there must exist a phase transition point $\mu = \Lambda_{EW}$ at which the additive renormalized Higgs mass parameter will approach to vanish for a proper UV characteristic energy scale $M_c$. Explicitly, we have 
\be
& & m_H^2(M_c/\Lambda_{EW}) =0,  \quad \Lambda_{EW} \simeq \sqrt{M_c^2 - \frac{(4\pi)^2  \tilde{m}_H^2(M_c/\Lambda_{EW})}{6y_H^2(M_c/\Lambda_{EW})}},
\ee
which shows that the large bare Higgs mass at the characteristic energy scale $M_c$ gets to become smaller and vanishing at 
a low energy phase transition point $\Lambda_{EW}$ through quadratic running. 

It becomes manifest that below such a phase transition point, i.e.,  $\mu < \Lambda_{EW}$, the additive renormalized Higgs mass parameter changes the sign. As a consequence, the Higgs potential gets unstable and the electroweak symmetry will be broken down spontaneously.  To be more explicit, we express the Higgs sector defined below the IR scale $\Lambda_{EW}$ as follows
\begin{equation}
\mathcal{L}_H=(D_{\mu}H)^{\dagger}(D^{\mu}H) + \mu_h^2H^{\dagger}H-\frac{\lambda_h}{4}(H^{\dagger}H)^2 -y_t\bar{Q}_Lt_R\tilde{H}+h.c., \quad \mu < \Lambda_{EW},
\end{equation}
with the definitions
\be
 \mu_h^2= \mu_h^2(\Lambda_{EW}/\mu)\equiv - m_{H}^{2}(M_c/\mu),\quad
 y_h^2 = y_h^2(\Lambda_{EW}/\mu) \equiv y_H^2(M_c/\mu),\quad  \lambda_h = \lambda_h(\Lambda_{EW}/\mu) \equiv \lambda_{H}(M_c/\mu),
\label{HiggsMassTerm}
\ee
where $\Lambda_{EW}$ is regarded as the electroweak symmetry breaking energy scale. 

When the electroweak symmetry gets broken down, the CP even neutral component of the Higgs doublet receives a nonzero evolving vacuum expectation value (eVEV), which is parametrized as the following form
\begin{equation}
H=\binom{H^+}{\frac{1}{\sqrt{2}}(\text{v}_h +h+i\chi)}, 
\label{HiggsBroken}
\end{equation}
where the Higgs field $h(x)$ corresponds to the quantized physical degree of freedom observed at the LHC. $\text{v}_h$ is the eVEV given by
\be
\text{v}^2_h=\text{v}^2_h(\Lambda_{EW}/\mu) \simeq \frac{4 \mu_h^2(\Lambda_{EW}/\mu) }{\lambda_h(\Lambda_{EW}/\mu)}.
\ee  
Such a symmetry breaking may be referred simply as the quantum electroweak symmetry breaking (QEWSB) as it is induced by the quadratic contributions of loop quantum effects. 

It is seen that the quantum loop quadratic contribution of top quark is crucial for the QEWSB as it mainly causes the additive renormalized Higgs mass parameter changing the sign. 
%%%%%%%%%%%%%%%%%%%%%%%%%%%%%%%%%%%%%%%% MODIFICATIONS %%%%%%%%%%%%%%%%%%%%%%%%%%%%%%%%%%%%%%%%%%%%%%

Before carrying out numerical calculations, let us remark further on the RGEs for $\mu<\Lambda_{EW}$. In this energy range, the RGEs of the dimensionless coupling constants such as, $g_1$, $g_2$, $g_3$, remain the same as those given before, while the RGE for the Higgs mass parameter $m^2_H(M_c/\mu)=-\mu^2_h(\Lambda_{EW}/\mu)$ with $\mu < \Lambda_{EW}$ is modified as all particles get masses after QEWSB. At one loop level, we have
\be
\mu_h^2(\Lambda_{EW}/\mu) & \simeq & \frac{6}{(4\pi)^2}\left(y^2_t-\frac{1}{4}\lambda_H-\frac{1}{8}g^2_1-\frac{3}{8}g^2_2\right)\left(\Lambda_{EW}^2-\mu^2 \right) - \frac{1}{(4\pi)^2}\left(  6 y^2_t m_t^2  - \frac{3}{2} \lambda_h  m_h^2 -\frac{9}{4}g^2_2 m_W^2  \right) \ln \frac{\Lambda_{EW}^2}{\mu^2}  \, ,\nonumber \\
& \simeq & \frac{6}{(4\pi)^2}\left(y^2_t-\frac{1}{4}\lambda_h-\frac{1}{8}g^2_1-\frac{3}{8}g^2_2\right)\left(\Lambda_{EW}^2-\mu^2 \right) - \frac{\mu_h^2}{(4\pi)^2}\left(  12 y^4_t/\lambda_h  - 3 \lambda_h  -\frac{9}{2}g^4_2/\lambda_h \right) \ln \frac{\Lambda_{EW}^2}{\mu^2} \, .
\ee
As the transition energy scale $\Lambda_{EW}$ is around TeV scale and the IR cut-off energy scale is set by the top quark mass $m_t \simeq 174$ GeV and the Higgs mass $m_h \simeq 125$ GeV, so that the above logarithmic correction is expected to be much smaller than the quadratic contribution.  The RGE can be read off as follows
\begin{equation}
\mu\frac{d \mu_h^2}{d \mu}=- \frac{12\mu^2}{(4\pi)^2} \bigg(y_t^2-\frac{1}{4}\lambda_H-\frac{1}{8}g_1^2-\frac{3}{8}g_2^2\bigg) + 
\frac{2\mu_h^2}{(4\pi)^2}\left(  12 y^4_t/\lambda_h  - 3 \lambda_H  -\frac{9}{2}g^4_2/\lambda_h \right).
\label{HiggsMassMod}
\end{equation}
which differs from Eq.~\eqref{QDHiggsMassRGE} for the logarithmic contributions. This is due to the fact that above $\Lambda_{EW}$ all the SM particles are massless except the Higgs boson, and thus there is no logarithmic contribution to the Higgs mass parameter except from the Higgs loop, which can be easily checked by carrying out the one-loop calculations explicitly. Below $\Lambda_{EW}$, the electroweak symmetry gets broken down spontaneously due to quantum loop effects of quadratic contributions, thus the SM particles such as the top quark and weak bosons start to gain nonzero masses through QEWSB. Then the top quark loops and weak boson loops, besides the Higgs loops, lead the logarithmic contributions to the Higgs mass parameter $\mu_h^2$.

In the numerical calculations, we shall take the RGEs \eqref{HiggsMassMod}, \eqref{gaugecouplingRGE}, \eqref{topyukawaRGE}, \eqref{higgscouplingRGE} with the boundary conditions at the top quark mass $\mu = m_t$, and then drive $\mu$ up to the energy scale $\mu =\Lambda_{EW}$ at which the additive renormalized Higgs mass parameter approaches to zero, i.e., $\mu_{h} ^2= -m_H^2(M_c/\Lambda_{EW}) = 0$. In other words, at $\mu =\Lambda_{EW}$ the electroweak symmetry gets restored (from the IR-to-UV viewpoint). 

For the boundary conditions at the top mass $m_t=173.34$ GeV, we take the latest extracted results presented in Ref.~\cite{buttazzo2013}
\be
& &  g_1(M_c/m_t)=0.35830,\quad g_2(M_c/m_t)=0.64779, \quad g_3(M_c/m_t)=1.1666, \nonumber \\
& & y_t(M_c/m_t)=0.93690,\quad \lambda_h(M_c/m_t)=0.50416,\quad \mu_h^2(M_c/m_t)=8652.7\ \text{GeV}^2.
\ee
Here we only take the central values as their errors without impacting on our final conclusion. Taking the above input as the boundary conditions for the RGEs \eqref{HiggsMassMod}, \eqref{gaugecouplingRGE}-\eqref{higgscouplingRGE}, we find the electroweak symmetry breaking scale $\Lambda_{EW}$ to be
\be
 \Lambda_{EW} \simeq 760\,  \mbox{GeV} \, .
\ee
The higher order corrections are generally suppressed due to the additional powers of coupling constants and $1/(4\pi)^2$ factor, and thus they play a less important role.

The additive renormalized Higgs mass parameter $m_H^2(M_c/\mu)$ is shown in Fig.~1.
%%%%%%%%%%%%%%%%%%%%%%%%%%%%%% FIGURE 1%%%%%%%%%%%%%%%%%%%%%%%%%%%%%%%%%%%%%
\begin{figure}[H]
\centering
\includegraphics[scale=0.3]{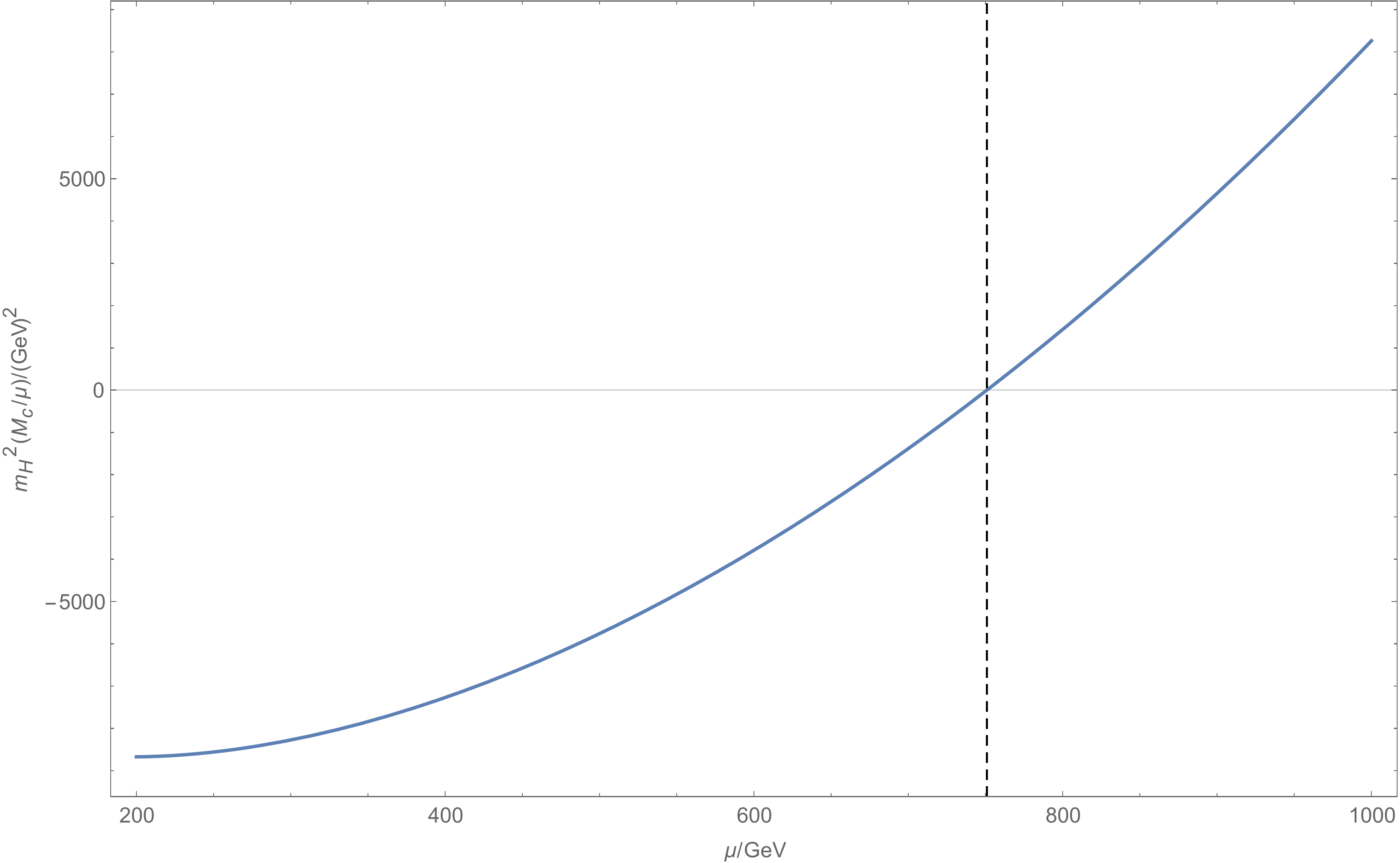}
\caption{The additive renormalized Higgs mass parameter $m_H^2(M_c/\mu)$ as the function of scale $\mu$ with $\mu=200 \sim 1000$ GeV.} 
\end{figure}
%%%%%%%%%%%%%%%%%%%%%%%%%%%%%%%%%%%%%%%%%%%%%%%%%%%%%%%%%%%%%%%%%%%%%%%%%

Before proceeding, let us remark on the gauge dependence about the vacuum expectation value of Higgs potential in the gauge theory, which is mainly caused via the gauge dependence of the renormalized Higgs field. Thus the Higgs VEV is intrinsically gauge dependent and cannot be treated as a physical observable in the gauge field theory. In much the same way, the eVEV defined in QEWSB is in general also gauge dependent. However, the gauge dependence of the eVEV shall in principle not affect the present analyses and conclusions. The reason is that the explicit form of eVEV could in general be changed when taking different gauge fixing conditions, while the pole masses and running parameters, such as the Higgs mass parameter $\mu_h^2$, which are concerned in the  present analysis, should in principle be gauge independent, thus the QEWSB mechanism and the resulting electroweak symmetry breaking scale $\Lambda_{EW}$ must be gauge independent. For more detailed discussions on the gauge dependence of the Higgs VEV, it is referred to the recent papers~\cite{Nielsen:2014spa, Andreassen:2014eha, Andreassen:2014gha} and references therein.

%%%%%%%%%%%%%%%%%%%%%%%%%%%%%%%%%%%%%%%%%%% MODIFICATIONS %%%%%%%%%%%%%%%%%%%%%%%%%%%%%%%%%%%%%%%%%%%%

When the sliding energy scale passes through the top mass threshold, i.e., $\mu \leq m_t$, the top quark will decouple effectively from the theory according to the usual assumption of effective field theory. This decoupling effect leaves a pattern on the later-on additive renormalized Higgs mass parameter as follows for $\mu \leq m_t$
\be
\mu_h^2(\Lambda_{EW}/\mu) & \simeq & \mu_h^2(\Lambda_{EW}/m_t)- \frac{3}{(4\pi)^2}\left(\frac{1}{2}\lambda_h+\frac{1}{4}g^2_1+\frac{3}{4}g^2_2\right)(m_t^2-\mu^2) + \frac{1}{(4\pi)^2}\left(  \frac{3}{2} \lambda_h  m_h^2 + \frac{9}{4}g^2_2 m_W^2  \right) \ln \frac{m_t^2}{\mu^2}  \nonumber \\
& \simeq &   \mu_h^2(\Lambda_{EW}/m_t) \left[ \, 1 +  \frac{1}{(4\pi)^2}\left( 3 \lambda_h  + \frac{9}{2}g^4_2/\lambda_h \right) \ln \frac{m_t^2}{\mu^2} \, \right] - \frac{3}{(4\pi)^2}\left(\frac{1}{2}\lambda_h+\frac{1}{4}g^2_1+\frac{3}{4}g^2_2\right)(m_t^2-\mu^2) 
 \ee  
which shows that the decoupling of the top quark leads the Higgs mass parameter $\mu_h^2$ to be decreased or $m_H^2(M_c/\mu)$ to be less negative. As a consequence, the eVEV $\text{v}_h^2(\Lambda_{EW}/\mu)$ gets a bit smaller. While such a decrease only happens in a small energy range as the sliding scale immediately moves down to the Higgs mass threshold $\mu = m_h= 125.15$ GeV \cite{buttazzo2013}. The actual change of the eVEV $\Delta \text{v}_h = \text{v}_h(\Lambda_{EW}/M_h)-\text{v}_h(\Lambda_{EW}/M_t)$ is tiny, numerically it is decreased only by about $1\%$.

When $\mu$ goes down further to be less than the Higgs mass $m_h$, the Higgs boson itself also decouples in much the same way as the top quark, which eventually freezes the renormalization of the Higgs mass parameter $\mu_h^2(\Lambda_{EW}/\mu) = -m_H^2(M_c/\mu)$ and the eVEV $\text{v}_h^2(\Lambda_{EW}/\mu)$. The numerical value of  the eVEV $\text{v}_h(\Lambda_{EW}/\mu)$ gets fixed and remains the same down to $\mu \simeq 0$, i.e., 
\be
 \text{v} \equiv \text{v}_h(\Lambda_{EW}/m_h)=\text{v}_{EW} \simeq 246 \mbox{ GeV}.
 \ee
The property of the additive renormalized Higgs mass parameter is illustrated in Fig.~2, where the anomalous changing between $\mu =m_h$ and $\mu = m_t $ is exaggerated. 
%%%%%%%%%%%%%%%%%%%%%%%%%%%%%% FIGURE 2%%%%%%%%%%%%%%%%%%%%%%%%%%%%%%%%%%%%%
\begin{figure}[H]
 \centering
\includegraphics[scale=0.5]{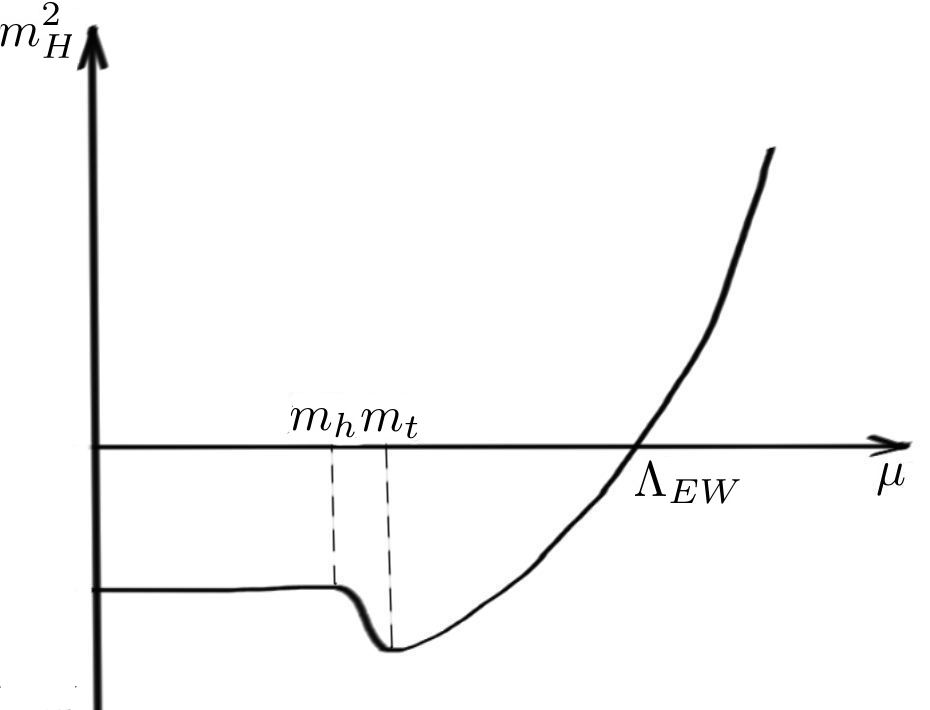}
 \label{HiggsMassRunning}
\caption{A qualitative sketch for the additive renormalized Higgs mass parameter $m_H^2(M_c/\mu)$ as the function of scale $\mu$. The electroweak symmetry gets broken down spontaneously at $\mu = \Lambda_{EW}\simeq 760$ GeV.}
 \end{figure}
%%%%%%%%%%%%%%%%%%%%%%%%%%%%%%%%%%%%%%%%%%%%%%%%%%%%%%%%%%%%%%%%%%%%%%%%%
For the logarithmic running with the multiplicative renormalization, the resulting Lagrangian can be shown to be scale-independence due to the cancellations of $\mu$ dependence among the coupling/mass renormalization and quantum fields renormalizations. Note that in applying for the renormalization scheme, a basic cut-off energy scale or a cut-off energy scale at a typical mass of particle is implicitly introduced to keep full control of the scale-independence. For instance, in QCD the basic QCD scale $\Lambda_{QCD} \simeq 300 $ MeV is used to have full control of the scale-independence of the strong interactions. For the quadratic running of the Higgs mass parameter, it is attributed to the additive renormalization, the scale-independence of the Higgs potential is realized simply by the subtraction between the additive renormalized Higgs mass and the running Higgs mass itself. For a detailed treatment up to the two-loop level, it is referred to Ref.~\cite{huangwu2012}.  In the present considerations, one can apply the characteristic energy scale $M_c$ and the phase transition energy scale $\Lambda_{EW} \simeq 760 $ GeV as well as the fixed eVEV $\text{v}_{EW} \simeq 246 $ GeV and masses of top quark and Higgs boson as the basic cut-off energy scales to make full control of the scale-independence of the Higgs potential. To be more explicit, we have
\be
& & m_H^2(M_c/\Lambda_{EW}) = m_H^2(M_c/\mu) + \mu_H^2(\mu/\Lambda_{EW}) =0, \nonumber \\
& & m_H^2(M_c/\mu)\simeq \tilde{m}_H^2(M_c/\mu)-\frac{6}{(4\pi)^2} y_H^2 \left(M_c^2-\mu^2\right), \nonumber \\
& & \mu_H^2(\mu/\Lambda_{EW})\simeq - \frac{3}{2(4\pi)^2} \lambda_H m_H^2 \text{ln}\frac{\mu^2}{\Lambda_{EW}^2}  -\frac{6}{(4\pi)^2} y_H^2 \left(\mu^2- \Lambda_{EW}^2 \right)\, .
\ee
As it turns out that for $\mu<\Lambda_{EW}$ the logarithmic contributions in the Higgs mass parameter are subdominant in comparison with the quadratic contributions, for simplicity of discussions, we shall only consider the quadratic terms as the inclusion of the logarithmic contributions is often straightforward. The $\mu$ independence of the Higgs mass parameter is manifest
\be
& & \mu_h^2(\Lambda_{EW}/m_h) = \mu_h^2(\Lambda_{EW}/m_t) +  \mu_h^2(m_t/\mu)  + \mu_h^2(\mu/m_h ), \nonumber \\ 
& & \mu_h^2(\Lambda_{EW}/m_t) \simeq \frac{6 y_h^2}{(4\pi)^2}\left(\Lambda_{EW}^2-m_t^2\right), \nonumber \\
& & \mu_h^2(m_t/\mu) \simeq - \frac{3}{(4\pi)^2}\left(\frac{1}{2}\lambda_h+\frac{1}{4}g^2_1+\frac{3}{4}g^2_2\right)(m_t^2-\mu^2), \nonumber \\
& & \mu_h^2(\mu/m_h) \simeq - \frac{3}{(4\pi)^2}\left(\frac{1}{2}\lambda_h+\frac{1}{4}g^2_1+\frac{3}{4}g^2_2\right)(\mu^2-m_h^2), 
\ee
where the $\mu$ dependence explicitly cancels each other.

In all the above analyses, we have assumed that the SM holds up to the UV characteristic energy scale $M_c$. Let us now study the possible value of $M_c$ in the SM with QEWSB mechanism. When the scale $\mu$ is above the QEWSB scale $\Lambda_{EW}$, the additive renormalized Higgs mass parameter $m_H^2(M_c/\mu)$ satisfies
\begin{equation}
m_H^2(M_c/\mu)\simeq \frac{6y_H^2(M_c/\mu) }{(4\pi)^2} (\mu^2- \Lambda_{EW}^2), \quad \mu > \Lambda_{EW}.
\end{equation}
For $\mu =M_c \gg \Lambda_{EW} \sim 760$ GeV, the bare Higgs mass is approximately given by 
\begin{equation}
m_H^2 \simeq \frac{6 y_H^2 }{(4\pi)^2} (M_c^2 -\Lambda_{EW}^2) \simeq \frac{6 y_H^2}{(4\pi)^2} M_c^2,
\label{BareHiggsMass}
\end{equation}
where the UV characteristic energy scale can in principle be taken to be the Planck scale $M_c \sim M_{Pl}$ or even a higher energy scale. On the other hand, it has been extensively discussed that the renormalized Higgs coupling $\lambda_H(M_c/\mu)$ approaches to zero at high energy scale. As mentioned before, a gauge independent extraction of the associated instability scale $\Lambda_c$ is around $10^{12}$ GeV \cite{Andreassen:2014gha}.  It implies that there must have some new physics beyond the SM in order to run the energy scale up to the grand unification scale or Planck scale $\mu \sim M_{Pl}$. 

Before going further, we would like to address that the above analyses are regularization scheme independent. In general, it has been shown\cite{wu2003,wu2014} that as long as the tensor-type and scalar-type irreducible loop integrals satisfy the consistent conditions of gauge invariance, one is able to arrive at  the right results as the scalar-type irreducible loop integrals can be evaluated by using any regularization schemes. The development of LORE method is to show explicitly that there does exist in principle a symmetry-preserving and infinity-free regularization method, which leads to the consistency conditions without modifying the original theory and changing the divergent structure of origin theory, especially the quadratic contributions of QFTs. It is well-known that the naive cut-off regularization scheme is not applicable when the Feynman loop integrals involve tensor-type irreducible loop integrals as it spoils the consistency conditions. 

Let us now make another interesting issue. That is whether the QEWSB mechanism in the SM is consistent with the expanding universe for $\mu \leq M_c$. More concretely, it is important to ensure that the Higgs boson with a renormalized mass is stable during the thermal history of our universe. It is widely believed that after the reheating the universe enters into the radiation dominant phase and starts the hot big bang. Then, the temperature $T$ of the cosmic plasma provides an effective measure of the typical energy scale for the particle physics processes taking place at that time. For $T> \Lambda_{EW}$, the SM lives in the electroweak symmetry phase and all the SM particles except the Higgs boson are massless. It is crucial to make sure that the Higgs boson does not decouple from the cosmic plasma too early, even before the QEWSB. Otherwise, the QEWSB mechanism does not work and all the SM particles cannot gain their mass from the electroweak symmetry breaking. The SM Higgs boson couples to the top quark most strongly and the coupling constant $y_t$ is roughly order one. For $T> \Lambda_{EW}$, the Higgs boson can decay into two massless top quarks and vice versa. The rate of this interaction is roughly $\Gamma \sim T$. Given the Hubble rate $H\sim T^2/M_{Pl}$, we have
\begin{equation}
\Gamma/H \sim M_{Pl}/T >1, \quad \text{for}\ \Lambda_{EW} < T < M_{Pl},
\end{equation}
which shows that the SM Higgs boson can always stay equilibrium with the cosmic plasma before the QEWSB. 

In conclusion,  we have demonstrated the QEWSB mechanism in the SM based on two postulations that (i) the Higgs boson has a large bare mass $m_H\gg m_h \simeq 125$ GeV at the UV characteristic energy scale $M_c$ which defines the SM in the UV region, and (ii) the quadratic contributions of loop effects are physically meaningful. It has been shown that the loop quadratic contributions in the SM can cause the additive renormalized Higgs mass parameter $m_H^{2}(M_c/\mu)$ to change the sign below the transition scale $\mu < \Lambda_{EW} \simeq 760$ GeV, which generates the spontaneous electroweak symmetry breaking. This is analogous to the chiral dynamical model of the low energy QCD \cite{daiwu2004} where the chiral symmetry breaking was shown to be induced dynamically by the loop quadratic contributions of light quarks with the scalar mesons regarded as the composite-Higgs particles. Unlike the chiral dynamical model of the low energy QCD,  the dominant loop quadratic contributions needed for the QEWSB arise from the heavy top quark. Although the calculations and physical interpretations have been made by using the LORE scheme in this article, the conclusions should be general as the analyses and results are regularization scheme independent and mainly based on two postulations mentioned above. The main reason for adopting the LORE scheme is that it is applicable for the Wilsonian effective field theory and preserves both the symmetry properties (e.g., gauge symmetry) and UV divergence structures (e.g., the quadratic and logarithmic divergences) in a manifest way in comparison with other schemes such as the naive UV cutoff and the dimensional regularization. Considering the fact that the energy scale of the hard scattering processes at the LHC has already reached the TeV scale, which is greater than the electroweak symmetry breaking scale $\Lambda_{EW} \simeq 760$ GeV obtained based on the QEWSB mechanism, it would be interesting to study the possible physics effects around the electroweak symmetry breaking scale $\Lambda_{EW} \simeq 760$ GeV and test the QEWSB mechanism at the LHC and the future Great Collider, which can be significantly important for a better understanding of both the hierarchy problem and naturalness problem. 

\section*{Acknowledgements}

The authors would like to thank Z. Liu for useful discussions. This work is supported in part by the National Nature Science Foundation of China (NSFC) under Grants No.~10975170, No.~10905084, No.~10821504; and the Project of Knowledge Innovation Program (PKIP) of the Chinese Academy of Science. This work is supported in part by the CAS Center for Excellence in Particle Physics (CCEPP).

%% The Appendices part is started with the command \appendix;
%% appendix sections are then done as normal sections
%% \appendix

%% \section{}
%% \label{}

%% If you have bibdatabase file and want bibtex to generate the
%% bibitems, please use
%%
%%  \bibliographystyle{elsarticle-num} 
%%  \bibliography{<your bibdatabase>}

\begin{thebibliography}{00}

%% \bibitem{label}
%% Text of bibliographic item

\bibitem{higgs2012a} G.~Aad et al.~[ATLAS Collaboration], Phys.~Lett.~B716, 1 (2012). 
\bibitem{higgs2012c} S.~Chatrchyan et al.~[CMS Collaboration], Phys.~Lett.~B716, 30 (2012).
\bibitem{Gross:1973id}
  D.~J.~Gross and F.~Wilczek,
  %``ULTRAVIOLET BEHAVIOR OF NON-ABELIAN GAUGE THEORIES,''
  Phys.~Rev.~Lett.~30, 1343 (1973).
\bibitem{Politzer:1973fx}
  H.~D.~Politzer,
  %``RELIABLE PERTURBATIVE RESULTS FOR STRONG INTERACTIONS?,''
  Phys.~Rev.~Lett.~30, 1346 (1973).
 \bibitem{RW} S.~P.~Robinson and F.~Wilczek, Phys.~Rev.~Lett.~96,
231601 (2006).
\bibitem{TW}
  Y.~Tang and Y.~L.~Wu,
%"Gravitational Contributions to Gauge Green's Functions and Asymptotic Free Power-Law Running of Gauge Coupling."
JHEP 1111, 073 (2011).
\bibitem{wilson1974} K.~Wilson and J.~Kogut, Phys.~Rept.~12, 75 (1974).
\bibitem{polchinski1984} J.~Polchinski, Nucl.~Phys.~B231, 269 (1984).

\bibitem{martin2011} See e.g., S.~Martin, hep-ph/9709356.
\bibitem{cheng2010} See e.g., H.C.~Cheng, arXiv:1003.1162; T.~Gherghetta, arXiv:1008.2570.
\bibitem{contino2010}  See e.g., R.~Contino, arXiv:1005.4269.
\bibitem{arkani-hamed2005} See e.g., N.~Arkani-Hamed and S.~Dimopoulos, JHEP 0506, 073 (2005).
\bibitem{gherghetta1994} T.~Gherghetta, Phys.~Rev.~D50, 5985 (1994).
\bibitem{daiwu2004} Y.B.~Dai and Y.L.~Wu, Eur.~Phys.~J.~C39, s1 (2005); hep-ph/0304075.
\bibitem{BBG1987} W.A.~Bardeen, A.J.~Buras, and J.-M.~Gerard, Nucl.~Phys.~B293, 787 (1987) ; Phys.~Lett.~B192, 138 (1987) ; 211, 343 (1988) .
\bibitem{HPSW1992} J.~Heinrich, E.A.~Paschos, J.-M.~Schwarz, and Y.L.~Wu, Phys.~Lett.~B279, 140 (1992).
\bibitem{wu2001} Y.L.~Wu, Phys.~Rev.~D64, 016001 (2001).

\bibitem{hamada2012} Y.~Hamada, H.~Kawai and K.~Oda, Phys.~Rev.~D87, 053009 (2013); Phys. Rev. D89, 059901 (2014).
\bibitem{hamada2013} Y.~Hamada, H.~Kawai and K.~Oda, Prog.~Theor.~Exp.~Phys.~2014, 023B02 (2014).
\bibitem{hamada2014} Y.~Hamada, H.~ Kawai, K.~Oda and S.C.~Park, Phys.~Rev.~Lett.~112, 241301 (2014).
\bibitem{chankowski2014} P.H. Chankowski et al., arXiv: 1404.0548.
\bibitem{bardeen1995} W.A.~Bardeen, Report No.~FERMILAB-CONF-95-391-T.

\bibitem{wu2003} Y.L.~Wu, Int.~J.~Mod.~Phys.~A18, 5363 (2003).
\bibitem{cuitangwu2009} J.W.~Cui, Y.~Tang and Y.L.~Wu, Phys.~Rev.~D79, 125008 (2009). 
\bibitem{huangwu2012} D.~Huang and Y.L.~Wu, Eur.~Phys.~J.~C72, 2066 (2012).
\bibitem{HLW2013} D.~Huang, L.F.~Li, Y.L. Wu, Eur.~Phys.~J.~C73, 2353 (2013).
\bibitem{wu2014} Y.L.~Wu, Invited talk, Proceedings of the Conference in Honour of the 90th Birthday of Freeman Dyson, June 26, (2014), World Scientific Publishing Co Pte Ltd, edited by K.K. Phua, L. C. Kwek, N. P. Chang; 
\\ Int.~J.~Mod.~Phys.~A29, 1430007 (2014).

\bibitem{veltman1981} M.~Veltman, Acta Phys.~Polon.~B12, 437 (1981).
\bibitem{alsarhi1992} M.S.~Al-sarhi, I.~Jack and D.R.T. Jones, Z.~Phys.~C55, 283 (1992).
\bibitem{buttazzo2013} D.~Buttazzo et al., JHEP 1312, 089 (2013).

%\cite{Nielsen:2014spa}
\bibitem{Nielsen:2014spa} 
  N.~K.~Nielsen,
  %``Removing the gauge parameter dependence of the effective potential by a field redefinition,''
  Phys.~Rev.~D90, 036008 (2014).
  %[arXiv:1406.0788 [hep-ph]].
  %%CITATION = ARXIV:1406.0788;%%
  %8 citations counted in INSPIRE as of 31 Mar 2015

%\cite{Andreassen:2014eha}
\bibitem{Andreassen:2014eha} 
  A.~Andreassen, W.~Frost and M.~D.~Schwartz,
  %``Consistent Use of Effective Potentials,''
  Phys.~Rev.~D91, 016009 (2015).
 % [arXiv:1408.0287 [hep-ph]].
  %%CITATION = ARXIV:1408.0287;%%
  %5 citations counted in INSPIRE as of 31 Mar 2015
  
%\cite{Andreassen:2014gha}
\bibitem{Andreassen:2014gha} 
  A.~Andreassen, W.~Frost and M.~D.~Schwartz,
  %``Consistent Use of the Standard Model Effective Potential,''
  Phys.~Rev.~Lett.~113, 241801 (2014).
  %[arXiv:1408.0292 [hep-ph]].
  %%CITATION = ARXIV:1408.0292;%%
  %11 citations counted in INSPIRE as of 31 Mar 2015




\end{thebibliography}

%% else use the following coding to input the bibitems directly in the
%% TeX file.

\end{document}